\documentclass[english,reprint, citeautoscript, aip, jcp]{revtex4-1}
\usepackage[T1]{fontenc}
\usepackage[utf8]{inputenc}
\usepackage{verbatim}
\usepackage{calc}
\usepackage{amsmath}
\usepackage{graphicx}
\PassOptionsToPackage{normalem}{ulem}
\usepackage{ulem}

\makeatletter

\providecommand{\tabularnewline}{\\}

\usepackage{algorithm,algpseudocode}

\usepackage{color}

\makeatother

\usepackage{babel}
\begin{document}
\title{Simple and efficient algorithms for training machine learning potentials
to force data}
\author{Justin S. Smith}
\affiliation{Theoretical Division and CNLS, Los Alamos National Laboratory, Los
Alamos, New Mexico 87545, USA}
\author{Nicholas Lubbers}
\affiliation{Computer, Computational, and Statistical Sciences Division, Los Alamos
National Laboratory, Los Alamos, New Mexico 87545, USA}
\author{Aidan P. Thompson}
\affiliation{Center for Computing Research, Sandia National Laboratories, Albuquerque,
New Mexico 87185, USA}
\author{Kipton Barros}
\email{kbarros@lanl.gov}

\affiliation{Theoretical Division and CNLS, Los Alamos National Laboratory, Los
Alamos, New Mexico 87545, USA}
\begin{abstract}
Machine learning models, trained on data from \emph{ab initio }quantum
simulations, are yielding molecular dynamics potentials with unprecedented
accuracy. One limiting factor is the quantity of available training
data, which can be expensive to obtain. A quantum simulation often
provides all atomic forces, in addition to the total energy of the
system. These forces provide much more information than the energy
alone\emph{. }It may appear that training a model to this large quantity
of force data would introduce significant computational costs. Actually,
training to all available force data should only be a few times more
expensive than training to energies alone. Here, we present a new
algorithm for efficient force training, and benchmark its accuracy
by training to forces from real-world datasets for organic chemistry
and bulk aluminum.
\end{abstract}
\maketitle
\global\long\def\d{\mathrm{d}}%
\global\long\def\tr{\mathrm{\mathrm{tr}}\,}%

\section{Introduction}

Machine learning (ML) is driving the development of next-generation
interatomic potentials. By training the ML model to a large and diverse
dataset of \emph{ab initio }quantum simulations, one aims to build
a low-cost, high-fidelity emulator, valid over a wide space of atomic
configurations. For example, such ML potentials can be used as the
basis for large scale molecular dynamics simulations with unprecedented
accuracy~\cite{Smith20,Lu20}.

The reference\emph{ }data is generated by approximate solution to
the Schrödinger equation, typically using a tool such as density functional
theory (DFT). Under the Born Oppenheimer approximation, nuclei are
treated classically. Each reference calculation takes as input the
atomic configuration (nuclei positions and species) and outputs total
energy $E$. Often, once the total energy has been computed, forces
$\mathbf{f}_{i}=-\partial E/\partial\mathbf{r}_{i}$ for \emph{all
}atoms $i$ can be produced, at minimal additional cost. If possible
to acquire, these forces provide highly valuable training data for
the ML model. For a system with $N$ atoms, the collection of force
components comprise $3N$ times more data than the energy scalar.

The ML model predicts a potential energy surface $\hat{E}$ that is
hopefully a good approximation to the true energy $E$, even for configurations
outside the training set. To maximize generality, it is natural to
train the ML model such that its predicted energy $\hat{E}$ \emph{and}
forces $\hat{\mathbf{f}}_{i}=-\partial\hat{E}/\partial\mathbf{r}_{i}$
agree with reference data. It might appear that incorporating a large
quantity of force data into the training procedure would incur a large
increase in computational cost. Here, we show otherwise. In the context
of neural networks (or more generally, any method based on gradient-based
optimization of a loss function) one can train on energy \emph{and
}force data at a cost comparable to training on the energy data alone.
Readers familiar with ML frameworks (ML-F) such as PyTorch~\cite{Paszke19}
or TensorFlow~\cite{Abadi16} may recognize the above statement as
self-evident. The principle of reverse-mode automatic differentiation
(backpropagation) \emph{guarantees }that the gradient of a scalar
loss function can be efficiently calculated, \emph{independent }of
the number of gradient components~\cite{Griewank89}. The backpropagation
procedure effectively requires tracing backward through all computational
steps that were used to calculate the loss. An ML-F will automatically
execute this procedure to produce the full gradient.

Prototyping new ML codes inside an ML-F is an obvious choice. However,
there remain several reasons why certain \emph{production} ML codes
may wish to avoid use of an ML-F. An obvious one is that it can be
difficult to port existing codes into the constrained context of an
ML-F. Another reason may be memory constraints. By design, the ML-F
needs to track every computational operation, recording all associated
data, in order to backpropagate. It is often possible to design alternative
algorithms to calculate a gradient, for which memory requirements
are significantly reduced~\cite{Wang18b}. Finally, there is a question
of performance. Codes may wish to avoid an ML-F if they require types
of calculations that are not easily expressable in terms of high-level
tensor operations. Although next-generation ML-Fs such as JAX~\cite{Bradbury18,Schoenholz19}
and Zygote~\cite{Innes19} enable backproprogation through nearly
arbitrary Python or Julia code, costs arising from automated tracing
seem unavoidable.

Our main contribution is a simple algorithm to efficiently calculate
the full gradient of a loss function that directly incorporates force
data. This algorithm works with or \emph{without} an ML-F, and so
remains fully general. In particular, the method can be applied to
any existing neural network code that was designed to train to energy
data, including AENet~\cite{Artrith16}, N2P2~\cite{Singraber18,Singraber19},
ANI~\cite{Smith17}, and PINN~\cite{Pun19}.

In a practical implementation, the cost to evalute the loss gradient
may be about 3 times the cost to predict all forces, independent\emph{
}of the number of atoms and number of model parameters.

\section{Evaluating the loss gradient}

\subsection{Problem statement}

Our context is as follows: We seek to optimize (i.e, train) model
parameters $\theta$ such that the ML-predicted energy function $\hat{E}_{\theta}[\mathbf{r}]$
minimizes a loss function,
\[
\mathcal{L}=c_{1}\mathcal{L}_{\mathrm{energy}}+c_{2}\mathcal{L}_{\mathrm{force}}+\mathcal{L}_{\mathrm{reg}}.
\]
The terms
\begin{align}
\mathcal{L}_{\mathrm{energy}} & =\frac{1}{2}\langle(\hat{E}-E)^{2}\rangle\nonumber \\
\mathcal{L}_{\mathrm{force}} & =\frac{1}{2}\langle\sum_{i\in\mathrm{atoms}}|\hat{\mathbf{f}}_{i}-\mathbf{f}_{i}|^{2}\rangle,\label{eq:force_loss}
\end{align}
constrain the model predictions $\hat{E}$ and $\hat{\mathbf{f}}_{i}$
to match reference energy and force data. Angle brackets $\langle\cdot\rangle$
denote an average over the dataset. The final term $\mathcal{L}_{\mathrm{reg}}$
is a placeholder for various possible model regularization terms.
Coefficients $c_{1}$ and $c_{2}$ are fixed prior to training.

Optimization of model parameters $\theta$ typically involves some
variant of stochastic gradient descent, which requires the loss gradient
$\nabla_{\theta}\mathcal{L}$, or an approximation to it. A modern
neural network will typically have $10^{4}$ or more scalar components
in $\theta$ and, therefore, in $\nabla_{\theta}\mathcal{L}$. It
is essential to calculate this full loss gradient efficiently. One
can handle $\nabla_{\theta}\mathcal{L}_{\mathrm{energy}}$ with ordinary
backpropagation. Calculating $\nabla_{\theta}\mathcal{L}_{\mathrm{force}}$,
however, presents an interesting challenge.

To simplify notation, let us focus attention on a single data point,
e.g. a single DFT calculation. For a system with $N$ atoms, we define
\begin{equation}
L=\frac{1}{2}\sum_{i=1}^{N}|\hat{\mathbf{f}}_{i}-\mathbf{f}_{i}|^{2}.\label{eq:L}
\end{equation}
The target $\nabla_{\theta}\mathcal{L}_{\mathrm{force}}$ can be calculated
by averaging $\nabla_{\theta}L$ over all data points.

Our focus, then, is efficient calculation of $\nabla_{\theta}L$.
This appears difficult, because $L$ already incorporates derivatives
of the ML potential, via its dependence on
\begin{align}
\hat{\mathbf{f}}_{i} & =-\partial\hat{E}/\partial\mathbf{r}_{i}.
\end{align}
Naïve expansion indicates that $\nabla_{\theta}L$ involves all \emph{second
}derivatives $\partial^{2}\hat{E}/\partial\mathbf{r}_{i}\partial\theta_{j}$.
Fortunately, it is not necessary to evaluate all these components
individually. Below, we demonstrate two methods to efficiently calculate
$\nabla_{\theta}L$ at a cost comparable to the calculation of $L$
alone.

\subsection{Approach 1: Iterated backpropagation\label{subsec:Approach-1}}

As previously mentioned, ML frameworks (ML-F) such as PyTorch or TensorFlow
offer an efficient algorithm to calculate the full gradient $\nabla_{\theta}L$.
This calculation happens as follows. Using primitives provided by
the ML-F, the user writes a code to calculate energy $\hat{E}$ and
the loss $L$ in terms of $\hat{E}$ and $\hat{\mathbf{f}}$. The
ML-F will first execute the code to calculate $\hat{E}[\mathbf{r}]$,
tracing all dependencies on the atomic configuration $\mathbf{r}$
and the model parameters $\theta$. Operating backward on that trace,
the ML-F then efficiently calculates all forces $\hat{\mathbf{f}}$.
Once these forces are known, the ML-F can calculate the loss $L$.
Throughout the entire calculation of $L$ (including the backpropagation
phase to calculate $\hat{\mathbf{f}}$), \emph{tracing remains active}.
A second\emph{ }backpropagation step can then be performed, this time
to calculate the full gradient $\nabla_{\theta}L$. We emphasize that
the calculation of $\nabla_{\theta}L$ involves backpropagating through
the backpropagation step used to calculate $\hat{\mathbf{f}}$. In
other words, calculation of $\nabla_{\theta}L$ effectively requires
\emph{four }traversals\emph{ }of the computational graph to calculate
$\hat{E}[\mathbf{r}]$. Remarkably, these steps are completely automated
by the ML-F. Implementing iterated backpropagation without the help
of an ML-F would be a daunting task.

Many popular ML potentials have been written in an ML-F, for which
the iterated backpropagation strategy is a natural fit~\cite{Lubbers18,Schutt18,Wang18a,Yao18,Zubatyuk19,Unke19,Gao20,Gilmer20}.

\subsection{Approach 2: Directional derivative of the energy gradient\label{subsec:Approach-2}}

Here we demonstrate that it is possible to efficiently calculate the
full gradient $\nabla_{\theta}L$, even without the aid of an ML-F.
This algorithm should be applicable to any existing neural network
code. Our \emph{only assumption} is that subroutines are available
to efficiently calculate the energy $\hat{E}$, as well as its two
gradients, $\partial\hat{E}/\partial\mathbf{r}_{i}$ and $\nabla_{\theta}\hat{E}$.\footnote{For efficiency, these gradients should be evaluated using backpropagation.
Implementing a \emph{first }iteration of backpropagation manually
is not too difficult.}

The error in the force predictions are readily calculated,
\begin{equation}
\mathbf{g}_{i}[\mathbf{r}]=\hat{\mathbf{f}}_{i}[\mathbf{r}]-\mathbf{f}_{i}[\mathbf{r}].
\end{equation}
The loss gradient may then be written as
\begin{align}
\nabla_{\theta}L & =\nabla_{\theta}\frac{1}{2}\sum_{i=1}^{N}|\mathbf{g}_{i}|^{2}=\sum_{i=1}^{N}\mathbf{g}_{i}\cdot\nabla_{\theta}\hat{\mathbf{f}}_{i}.
\end{align}
In the second step we used the fact that $\mathbf{f}_{i}$ is ground
truth data, independent of model parameters $\theta$. Applying the
definition $\hat{\mathbf{f}}_{i}=-\partial\hat{E}/\partial\mathbf{r}_{i}$
and commuting derivatives yields
\begin{equation}
\nabla_{\theta}L=-\sum_{i=1}^{N}\mathbf{g}_{i}\cdot\frac{\partial}{\partial\mathbf{r}_{i}}(\nabla_{\theta}\hat{E}).
\end{equation}
Naïvely, one might consider evaluating $\nabla_{\theta}L$ by finite
differencing on all $N$ positions $\mathbf{r}_{i}$ individually.
There is a better algorithm, however, which avoids introducing a factor
of $N$ into the computational cost.

The idea is to interpret $\mathbf{g}=[\mathbf{g}_{1},\mathbf{g}_{2}\dots\mathbf{g}_{N}]$
as a $3N$-dimensional vector in the space of all atomic coordinates,
and$\frac{\partial}{\partial\mathbf{r}}=[\frac{\partial}{\partial\mathbf{r}_{1}},\frac{\partial}{\partial\mathbf{r}_{2}},\dots\frac{\partial}{\partial\mathbf{r}_{N}}]$
as the gradient vector in this space. In this language, the loss gradient
($\nabla_{\theta}L$) may be viewed as a \emph{directional derivative}
of the energy gradient $(\nabla_{\theta}\hat{E})$ along the direction
of force errors ($\mathbf{g}$). Central differencing gives,
\begin{equation}
\nabla_{\theta}L\approx\tilde{\nabla}_{\theta}L=-\frac{\nabla_{\theta}\hat{E}[\mathbf{r}_{+}]-\nabla_{\theta}\hat{E}[\mathbf{r}_{-}]}{2\eta},\label{eq:grad_L}
\end{equation}
where $\mathbf{r}_{+}$ and $\mathbf{r}_{-}$ denote new configurations
in which each atom is slightly perturbed,
\begin{equation}
\left(\mathbf{r}_{\pm}\right)_{i}=\mathbf{r}_{i}\pm\eta\mathbf{g}_{i}.\label{eq:r_pm}
\end{equation}
In Eq.~(\ref{eq:grad_L}), $\mathbf{r}_{+}$ and $\mathbf{r}_{-}$
are to be held fixed with respect to varations in $\theta$ (namely,
we \emph{impose} $\nabla_{\theta}\mathbf{r}_{\pm}=0)$. Models $\hat{E}$
are typically designed to be smooth, so Eq.~(\ref{eq:grad_L}) is
valid to order $\mathcal{O}(\eta^{2}).$ The ``small'' parameter
$\eta$ has units of length per force. Its selection will be discussed
below.

Combining the above results, our method can be summarized as follows:

\noindent\fbox{\begin{minipage}[t]{1\columnwidth - 2\fboxsep - 2\fboxrule}%
\textbf{Steps for efficient evaluation of loss gradient}
\begin{enumerate}
\item For a given atomic configuration $\mathbf{r}$, calculate all predicted
forces $\hat{\mathbf{f}}$, and associated force errors, $\mathbf{g}=\mathbf{\hat{f}}-\mathbf{f}$.
\item Generate slightly perturbed atomic configurations $\mathbf{r}_{\pm}=\mathbf{r}\pm\eta\mathbf{g}$
\item Evaluate the full energy gradient $\nabla_{\theta}\hat{E}$ at new
configurations $\mathbf{r}_{+}$ and $\mathbf{r}_{-}$.
\item Use central differences, Eq.~(\ref{eq:grad_L}), to approximate $\nabla_{\theta}L=\tilde{\nabla}_{\theta}L+\mathcal{O}(\eta^{2})$.
\end{enumerate}
\end{minipage}}

In total, this recipe requires calculating forces $\mathbf{\hat{f}}$
and two additional energy gradients, $\nabla_{\theta}\hat{E}[\mathbf{r}_{+}]$,
and $\nabla_{\theta}\hat{E}[\mathbf{r}_{-}]$. Compared to the method
of Sec.~\ref{subsec:Approach-1}, less memory is required because
here we avoid iterated\emph{ }backpropagation.

Equation~(\ref{eq:grad_L}) may be interpreted as an approximation
to Pearlmutter's algorithm for efficient multiplication by the Hessian~\cite{Pearlmutter94}.
In Pearlmutter's version, the $\eta\rightarrow0$ limit is taken,
avoiding numerical errors due to finite differencing. This can be
achieved using the method of \emph{forward mode }automatic differentiation~\cite{Griewank89}.
Specifically, the code to calculate $\nabla_{\theta}\hat{E}$ should
be transformed into one that operates on so-called dual numbers, which
are designed to track infinitesimal perturbations. Unlike reverse
mode autodiff, the forward mode version requires no tracing.

Existing neural network codes are unlikely to support dual numbers,
so we instead advocate the central difference approximation of Eq.~(\ref{eq:grad_L}).
The next section will indicate that numerical errors can be quite
small.

\section{Minimizing numerical error\label{sec:Minimizing}}

\subsection{Scaling the finite differencing parameter}

The finite differencing scheme of Eq.~(\ref{eq:grad_L}) requires
selection of a sufficiently small parameter $\eta$. Since $\eta$
actually carries dimensions, it is natural to factorize
\begin{equation}
\eta=\epsilon a_{0}/g_{0},\label{eq:eta}
\end{equation}
where $a_{0}$ is a characteristic length scale, and $g_{0}$ is a
characteristic scale associated with errors in the force predictions,
$\mathbf{g}$. The small dimensionless parameter $\epsilon$ should
be selected according to floating point round-off considerations,
as will be discussed below.

For simplicity, we fix $a_{0}=\mathrm{\mathring{A}}$. The characteristic
scale $g_{0}$ should vary according to the accuracy of the model's
force predictions, as applied to a particular system. A reasonable
choice is
\begin{equation}
g_{0}=\max_{i=1\dots N}|\mathbf{g}_{i}|,\label{eq:g0}
\end{equation}
selected on a per system basis.

\subsection{Two measures of error}

A direct error measure for the finite differencing scheme of Eq.~(\ref{eq:grad_L})
is, 
\begin{equation}
\mathrm{Err}[\tilde{\nabla}L]=\frac{|\tilde{\nabla}_{\theta}L-\nabla_{\theta}L|}{|\nabla_{\theta}L|}.
\end{equation}
The bars $|\cdot|$ denote an $L_{2}$ norm, to be taken over all
$\theta$ components, and all points in the dataset (e.g. all DFT
calculations).

Ideally, one would like to know how floating point round-off errors
contribute to $\mathrm{Err}[\tilde{\nabla}L]$. In applications, it
may be infeasible to calculate $\nabla_{\theta}L$ to full precision,
and we therefore will not know the true numerical error in $\tilde{\nabla}_{\theta}L$.
To circumvent this limitation, it will be useful to introduce a second
error measure that can be used as a proxy for $\mathrm{Err}[\tilde{\nabla}L]$.

Removing the gradient operator $\nabla_{\theta}$ from the right hand
side of Eq.~(\ref{eq:grad_L}) yields a new finite difference approximation,
\begin{equation}
\frac{\hat{E}[\mathbf{r}_{+}]-\hat{E}[\mathbf{r}_{-}]}{2\eta}\approx\sum_{i}\mathbf{g}_{i}\cdot\frac{\partial\hat{E}}{\partial\mathbf{r}_{i}},\label{eq:dE_deta}
\end{equation}
again valid to order $\mathcal{O}(\eta^{2})$. Using $L=\frac{1}{2}\sum_{i}\mathbf{g}_{i}\cdot(\hat{\mathbf{f}}_{i}-\mathbf{f}_{i})$,
we find that
\begin{equation}
L\approx\tilde{L}=-\frac{\hat{E}[\mathbf{r}_{+}]-\hat{E}[\mathbf{r}_{-}]}{4\eta}-\frac{1}{2}\sum_{i}\mathbf{g}_{i}\cdot\mathbf{f}_{i}.\label{eq:L_tilde}
\end{equation}
The suggests a new error measure
\begin{equation}
\mathrm{Err}[\tilde{L}]=\frac{|\tilde{L}-L|}{|L|},
\end{equation}
which should reflect $\mathrm{Err}[\tilde{\nabla}L]$, insofar as
the finite difference approximations Eqs.~(\ref{eq:grad_L}) and~(\ref{eq:L_tilde})
have comparable round-off errors. Below we present evidence to this
effect.

Because the reference loss $L$ is generally available, the error
measure $\mathrm{Err}[\tilde{L}]$ can be calculated to high precision
with existing codes.

\subsection{Empirical error measurements\label{subsec:empirical-error}}

Here we demonstrate a numerical procedure for selecting the dimensionless
parameter $\epsilon$, which fixes $\eta$ via Eq.~(\ref{eq:eta}).

Our intention is that the approximate loss gradient $\tilde{\nabla}_{\theta}L$
will ultimately be used to enable a gradient descent training procedure,
for which the ML model will have a highly nonlinear dependence on
its model parameters $\theta$. In this subsection, however, we consider
the simpler context of a linear regression model so that it becomes
possible to precisely evaluate the effects of floating point round-off
on $\mathrm{Err}[\tilde{\nabla}L]$. Local energy contributions will
be modeled as $\hat{E}=\sum_{j}\theta_{j}B_{j}$, where $\theta_{j}$
are fitting coefficients and $B_{j}$ serve as descriptors of each
local atomic environment. For concreteness, we select a SNAP potential
for tantalum, and use its corresponding dataset~\cite{Trott14,Thompson15}.
In SNAP, the descriptors $B_{j}$ are bispectrum coefficients. The
dataset consists of 362 different configurations, sampled from both
crystal and liquid phases~\cite{Thompson15}. System sizes in this
dataset range from 2 to 100 tantalum atoms. Reference energy and force
data were computed with DFT.

In the context of an ML training procedure, we must account for the
fact that the parameters $\theta$ will be rapidly evolving. Ideally,
$\tilde{\nabla}_{\theta}L$ should remain a good approximation to
$\nabla_{\theta}L$ for \emph{arbitrary }model parameters $\theta$.
Therefore, in addition to the trained SNAP potential, we also consider
an \emph{untrained} model, for which we randomize the model parameters
$\theta_{j}$ according to the Kaiming initialization procedure~\cite{He15}.

\begin{figure}
\includegraphics[width=0.9\columnwidth]{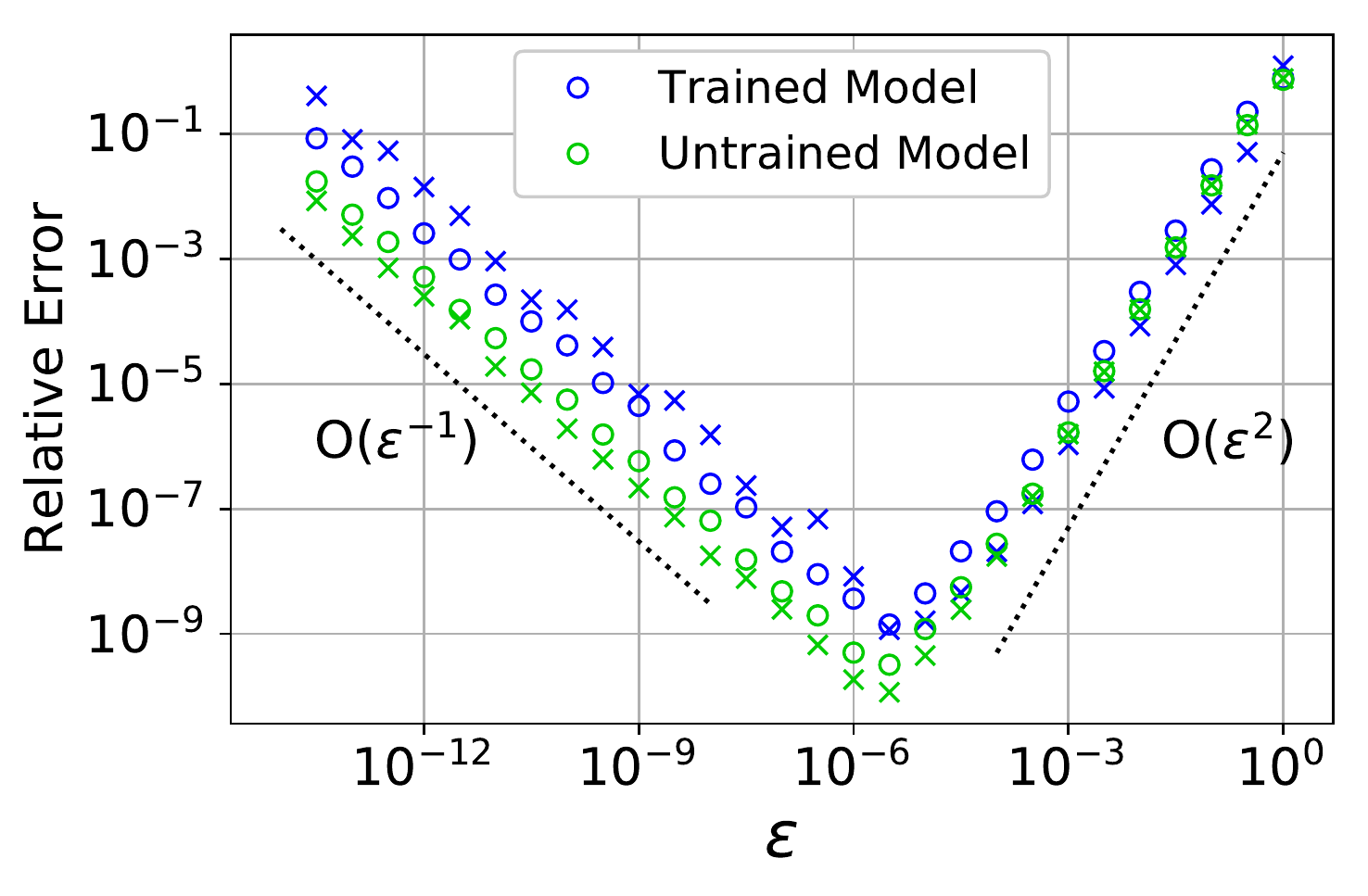}\caption{Relative error in the finite difference estimates of $\nabla_{\theta}L$
for trained and untrained SNAP potentials of tantalum. Circles denote
the true error $\mathrm{Err}[\tilde{\nabla}L]$, and crosses denote
its proxy $\mathrm{Err}[\tilde{L}]$. Central differencing errors
formally scale as $\mathcal{O}(\epsilon^{2})$ in the small parameter
$\epsilon$. Accounting for double-precision round-off errors, the
choice $\epsilon\approx10^{-5}$ yields the smallest errors for both
model types (under both error measures).\label{fig:err}}
\end{figure}

Figure~\ref{fig:err} shows empirical measurements of errors associated
with the approximation $\tilde{\nabla}_{\theta}L\approx\nabla_{\theta}L$,
for various values of the dimensionless finite differencing parameter
$\epsilon$. Importantly, a \emph{single }value, $\epsilon\approx10^{-5}$,
is observed to minimize the error for both trained and untrained models.
This optimal $\epsilon$ balances the $\mathcal{O}(\epsilon^{2})$
central differencing error with the floating point round-off error.
For this calculation, we used 64-bit (double-precision) floating point
accuracy, for which the 53 bit significand corresponds to approximately
$16$ digits\emph{ }of precision. Here, the proper selection of $\epsilon$
yields about 9 digits of accuracy in estimates $\tilde{\nabla}_{\theta}L$
of the loss gradient, $\nabla_{\theta}L$, which is more than sufficient
for neural network training.

Figure~\ref{fig:err} actually reports our \emph{two} measures of
error, namely $\mathrm{Err}[\tilde{\nabla}L]$ and $\mathrm{Err}[\tilde{L}]$.
Recall that the former represents the true\emph{ }error in the approximation
$\tilde{\nabla}_{\theta}L\approx\nabla_{\theta}L$, and the latter
is intended as a proxy for the true error. Our results indicate that,
indeed, $\mathrm{Err}[\tilde{\nabla}L]$ and $\mathrm{Err}[\tilde{L}]$
are of comparable scale. When moving to real-world neural network
codes, $\mathrm{Err}[\tilde{L}]$ will be easy to directly measure.
Our general recommendation is to select $\epsilon$ to minimize $\mathrm{Err}[\tilde{L}]$.
The results of Fig.~\ref{fig:err} indicate that this choice of $\epsilon$
will yield a quality approximation $\tilde{\nabla}_{\theta}L\approx\nabla_{\theta}L$,
even under significant variations to the model parameters $\theta$.

The scaling relations of Eqs.~(\ref{eq:eta}) and~(\ref{eq:g0})
are crucial for ensuring that $\epsilon$ is roughly invariant to
model quality. In particular, as the model improves, the typical force
errors $\mathbf{g}=\hat{\mathbf{f}}-\mathbf{f}$ decrease, and the
finite differencing parameter $\eta$ should \emph{increase}, so that
the characteristic atomic displacements $\mathbf{r}_{\pm}-\mathbf{r}=\pm\eta\mathbf{g}$
of Eq.~(\ref{eq:r_pm}) have a roughly invariant scale. Appendix~\ref{sec:more_errors}
demonstrates the importance of accounting for these scaling relationships.

\section{Benefits of force training}

Our initial motivation for developing the force training scheme of
Sec.~\ref{subsec:Approach-2} was to support ANAKIN-ME (ANI) models~\cite{Smith17}.
ANI is a variant of the Behler Parrinello neural network architecture~\cite{Behler07}.
The Neurochem implementation of ANI is written in highly optimized
C++/CUDA code~\cite{Smith20b}, and does not use an ML framework
such as PyTorch or TensorFlow. Applied to a recently developed aluminum
potential, NeuroChem can calculate 1,000 atomic forces in about 20
ms, running on a single modern GPU (Nvidia RTX 2080 Ti)~\cite{Smith20}.
For comparison, TorchANI is a recent reimplemenation of ANI in PyTorch,
designed for flexibility~\cite{Gao20}. TorchANI makes prototyping
new model variants much easier, but is up to an order of magnitude
slower than NeuroChem. Whereas TorchANI can use the iterated backpropagation
scheme of Sec.~\ref{subsec:Approach-1}, the optimized NeuroChem
implementation cannot. Fortunately, the method presented in Sec.~\ref{subsec:Approach-2}
allows NeuroChem to also train to force data in a very efficient manner.

We demonstrate the value of training to force data by benchmarking
on two real-world datasets. The first, ANI-1x, includes about 5M DFT
calculations on single organic molecules (elements C, H, N, and O,
with a mean molecule size of about 15 atoms), over a broad range of
conformations~\cite{Smith20a}. The second, ANI-Al, includes about
6,000 DFT calculations on bulk aluminum, in various melt and crystal
configurations, each containing about 100 to 200 atoms~\cite{Smith20}.
Both datasets were generated automatically using an active learning
procedure, which aims to maximize the diversity of the atomic configurations~\cite{Smith18}.
For each of the two datasets, we trained two ML potentials. The first
potential was trained to energy data only, and the second potential
was trained to both energy \emph{and }force data. We employ ensemble
averaging to reduce model variance; each ML potential actually consists
of eight ANI neural networks, differing only in the random initialization
of their weights prior to training. Model details and training procedures
are described in previous work~\cite{Smith18,Smith20}.

The force training scheme of \ref{subsec:Approach-2} requires selection
of a finite differencing parameter $\eta$ via the dimensionless number
$\epsilon$. The NeuroChem implementation uses a careful mix of 32
bit and 64 bit floating point precision, and the optimal choice of
$\epsilon$ would be difficult to guess \emph{a priori. }We selected
$\epsilon=10^{-3}$ to approximately minimize $\mathrm{Err}[\tilde{L}]$,
and found that this choice yields reasonable estimates of the loss
gradient $\nabla_{\theta}L$ throughout the training procedure.

\begin{table}
\begin{tabular}{lll}
\hline 
 & ANI-1x (chem.)~~~~~~~ & ANI-Al (alum.)\tabularnewline
\hline 
\multicolumn{3}{c}{\uline{Training on energy data only}}\tabularnewline
Energy RMSE~~~~~ & $1.48\pm0.01\,\frac{\mathrm{kcal}}{\mathrm{mol}}$ & $4.38\pm0.45\,\frac{\mathrm{meV}}{\mathrm{atom}}$\tabularnewline
Force RMSE & $4.12\pm0.02\,\frac{\mathrm{kcal}}{\mathrm{mol\,\mathring{A}}}$ & $0.39\pm0.05\,\frac{\mathrm{eV}}{\mathrm{\mathring{A}}}$\tabularnewline
\multicolumn{3}{c}{\uline{Training on energy }\emph{\uline{and}}\uline{ force
data}}\tabularnewline
Energy RMSE & $\mathbf{1.38}\pm0.01\,\frac{\mathrm{kcal}}{\mathrm{mol}}$ & $\mathbf{1.88}\pm0.2\,\frac{\mathrm{meV}}{\mathrm{atom}}$\tabularnewline
Force RMSE & $\mathbf{2.78}\pm0.014\,\frac{\mathrm{kcal}}{\mathrm{mol\,\mathring{A}}}$ & $\mathbf{0.045}\pm0.001\,\frac{\mathrm{eV}}{\mathrm{\mathring{A}}}$\tabularnewline
\end{tabular}\caption{Root-mean-squared-errors (RSME) for neural network energy and force
predictions. Models were trained to the ANI-1x and ANI-Al datasets
for organic chemistry and bulk aluminum, respectively.\label{tab:error}}
\end{table}

Table~\ref{fig:err} shows the resulting errors in energy and force
predictions, as measured on held-out test data. Because the natural
energy units vary according to domain, we use kcal/mol for the ANI-1x
dataset (organic chemistry) and eV for the ANI-Al data (bulk aluminum).

For ANI-1x, including force data into the training procedure reduces
error in the energy predictions by about 7\%, and in the force predictions
by about 33\%. For ANI-Al, we see a much more dramatic improvement:
energy and force errors are reduced by about 57\% and 88\%, respectively.
In other words, using force data in the training procedure can reduce
force prediction errors by almost a factor of 9.

The biggest difference between the ANI-1x and ANI-Al datasets is that
the latter contains DFT calculations for \emph{bulk} systems (100
to 200 aluminum atoms), whereas the former contains calculations for
single molecules only (each with about 15 atoms on average). Consequently,
in the ANI-Al dataset, far more information is carried by the force
data than the energy data.

\section{Conclusions}

Various works state or imply that training neural network potentials
to force data is challenging or expensive. Some studies even opt to
ignore forces, and train only to energies. Here, we have discussed
two algorithms that make training to force data simple and efficient.
With either algorithm, the computational cost of training to energy
\emph{and }force data is only a few times more expensive than the
cost of training to energy data alone, independent of system size
and model complexity. This is striking given that, for a bulk system,
the collection of all forces contains vastly more information than
does the energy alone.

In Sec.~\ref{subsec:Approach-1} we discussed the method of \emph{iterative
backprogation}. Using an ML framework such as PyTorch or TensorFlow,
force training can be handled almost automatically. One is free to
place arbitrary force-dependent terms into the loss function, and
gradients come ``for free.'' Under the hood, the ML framework traces
all intermediate values in the computational graph for calculating
the loss function, and will automatically backpropagate through this
graph to calculate the gradient of the loss function. We use the term
``iterated backpropagation'' to refer to the fact backpropagation
must happen twice (first to calculate forces and second to calculate
the gradient of the loss).

In Sec.~\ref{subsec:Approach-2} we presented a new method that involves
taking an appropriate \emph{directional derivative of the energy gradient}.
A primary motivation for the new method is that it does not require
the use of an ML framework; our method could be applied to \emph{any}
existing\emph{ }neural network code that was designed to train to
energy data. Compared to iterated backpropagation, the new method
requires only half the memory, because it avoids the second backpropagation
step. The directional derivative may be estimated with \emph{single
}central difference approximation of Eq.~(\ref{eq:grad_L}). The
numerical errors associated with finite differencing were found to
be negligible in practice. Alternatively, one could in principle retain
\emph{full }numerical precision in calculating the loss gradient if
the neural network code happens to support a generalization to dual
numbers~\cite{Griewank89,Pearlmutter94}.

The benefits of force training have been extensively demonstrated
in Ref.~\onlinecite{Cooper20}. Interestingly, that study treats
the loss function $L$ of Eq.~(\ref{eq:force_loss}) in a more approximate
way. Namely, the authors reframed the problem in terms of energy training
only, by effectively augmenting their dataset with small, random perturbations
to existing configurations. Here, in contrast, here we have shown
it possible to directly calculate the \emph{full }gradient $\nabla_{\theta}L$
at a cost only a few times greater than the cost to calculate $L$
itself, independent of the system size and the number of model parameters.

We have focused on ML models for which training involves some flavor
of gradient descent optimization. Kernel methods, such as Gaussian
process regression, are an alternative approach to ML potential development,
for which the model parameters are calculated via solution to a linear
system of equations~\cite{Bartok15,Rupp15}. Force training is important
for kernel models as well as for neural networks~\cite{Chmiela17,Christensen19}.
One might ask: Could the algorithms presented here also be of use
when training kernel models to large quantities of force data?
\begin{acknowledgments}
This work was partially supported supported by the Laboratory Directed
Research and Development (LDRD) program at LANL. N.~L. and A.~T.
acknowledge support from the Exascale Computing Project (17-SC-20-SC),
a collaborative effort of the U.S. Department of Energy Office of
Science and the National Nuclear Security Administration. K.~B. acknowledges
support from the center of Materials Theory as a part of the Computational
Materials Science (CMS) program, funded by the U.S. Department of
Energy, Office of Science, Basic Energy Sciences, Materials Sciences
and Engineering Division. Sandia National Laboratories is a multimission
laboratory managed and operated by National Technology \& Engineering
Solutions of Sandia, LLC, a wholly owned subsidiary of Honeywell International
Inc., for the U.S. Department of Energy’s National Nuclear Security
Administration under contract DE-NA0003525. This paper describes objective
technical results and analysis. Any subjective views or opinions that
might be expressed in the paper do not necessarily represent the views
of the U.S. Department of Energy or the United States Government.
\end{acknowledgments}

\appendix

\section{Importance of proper $\eta$ scaling\label{sec:more_errors}}

\begin{figure}
\includegraphics[width=0.9\columnwidth]{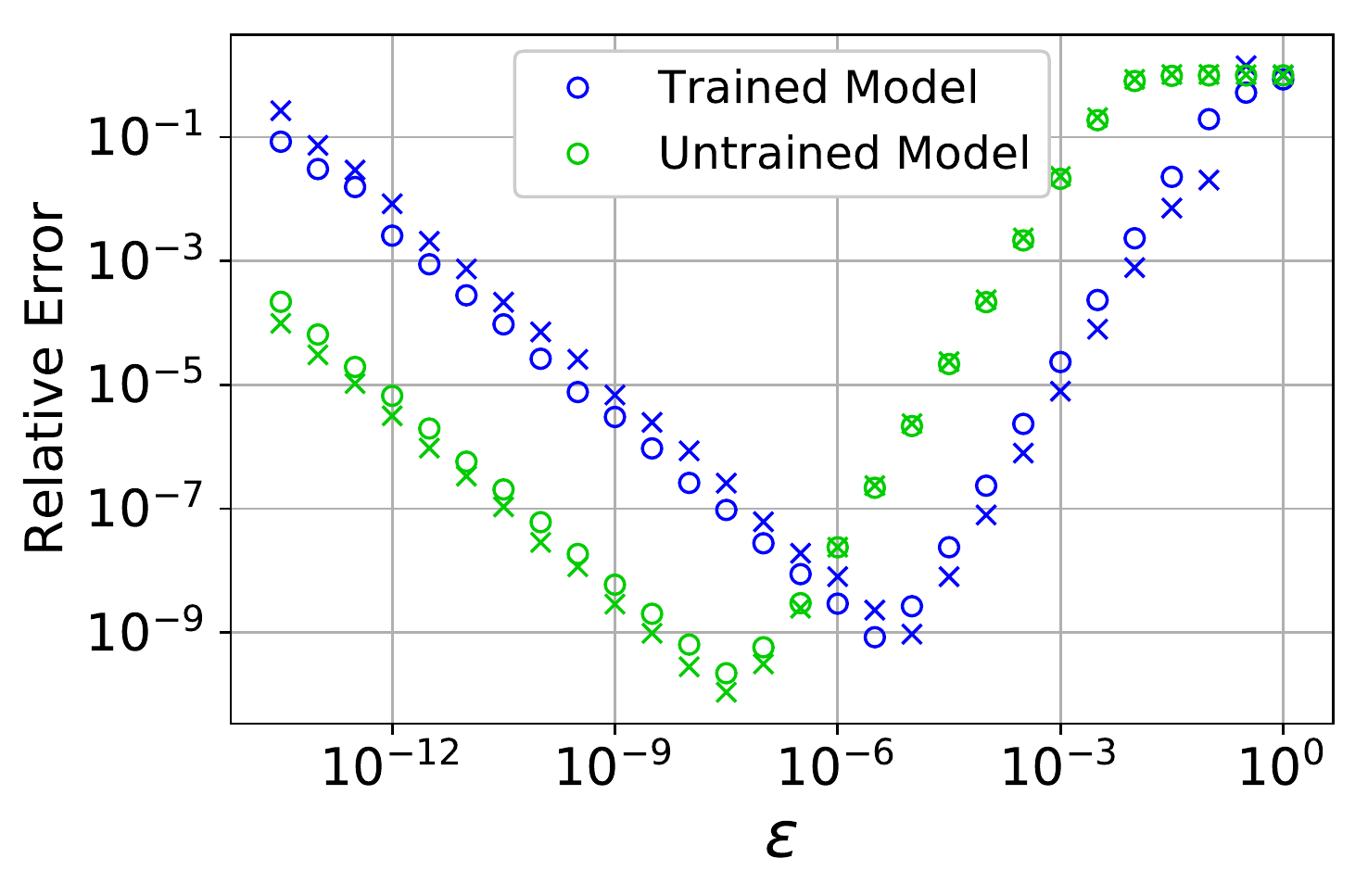}\caption{Errors in the finite difference estimates of $\nabla_{\theta}L$,
analogous to those of Fig.~(\ref{fig:err}), but here naïvely fixing
$g_{0}=\mathrm{eV/\mathring{A}}$, rather than using the definition
of Eq.~(\ref{eq:g0}). With this replacement, the optimal values
of $\epsilon$ now differ significantly between trained and untrained
models.\label{fig:err2}}
\end{figure}
Figure~\ref{fig:err} measured errors in the approximation $\tilde{\nabla}_{\theta}L\approx\nabla_{\theta}L$
for trained and untrained SNAP potentials. By scaling $\eta$ according
to Eqs.~(\ref{eq:eta}) and~(\ref{eq:g0}), we achieved a good approximator
$\tilde{\nabla}_{\theta}L$, valid for both trained and untrained
models, using a \emph{single} dimensionless parameter $\epsilon$.
The invariance of $\epsilon$ is important because one expects model
parameters $\theta$ to vary significantly over the course of an ML
training procedure.

Figure~\ref{fig:err2} illustrates the danger of naïvely fixing $g_{0}$
constant, rather than using Eq.~(\ref{eq:g0}). We observe that,
with $g_{0}$ fixed, the optimal value of $\epsilon$ can easily vary
by multiple orders of magnitude between trained and untrained models.

\bibliographystyle{apsrev4-1}
\bibliography{force_training}

\begin{thebibliography}{36}%
\makeatletter
\providecommand \@ifxundefined [1]{%
 \@ifx{#1\undefined}
}%
\providecommand \@ifnum [1]{%
 \ifnum #1\expandafter \@firstoftwo
 \else \expandafter \@secondoftwo
 \fi
}%
\providecommand \@ifx [1]{%
 \ifx #1\expandafter \@firstoftwo
 \else \expandafter \@secondoftwo
 \fi
}%
\providecommand \natexlab [1]{#1}%
\providecommand \enquote  [1]{``#1''}%
\providecommand \bibnamefont  [1]{#1}%
\providecommand \bibfnamefont [1]{#1}%
\providecommand \citenamefont [1]{#1}%
\providecommand \href@noop [0]{\@secondoftwo}%
\providecommand \href [0]{\begingroup \@sanitize@url \@href}%
\providecommand \@href[1]{\@@startlink{#1}\@@href}%
\providecommand \@@href[1]{\endgroup#1\@@endlink}%
\providecommand \@sanitize@url [0]{\catcode `\\12\catcode `\$12\catcode
  `\&12\catcode `\#12\catcode `\^12\catcode `\_12\catcode `\%12\relax}%
\providecommand \@@startlink[1]{}%
\providecommand \@@endlink[0]{}%
\providecommand \url  [0]{\begingroup\@sanitize@url \@url }%
\providecommand \@url [1]{\endgroup\@href {#1}{\urlprefix }}%
\providecommand \urlprefix  [0]{URL }%
\providecommand \Eprint [0]{\href }%
\providecommand \doibase [0]{http://dx.doi.org/}%
\providecommand \selectlanguage [0]{\@gobble}%
\providecommand \bibinfo  [0]{\@secondoftwo}%
\providecommand \bibfield  [0]{\@secondoftwo}%
\providecommand \translation [1]{[#1]}%
\providecommand \BibitemOpen [0]{}%
\providecommand \bibitemStop [0]{}%
\providecommand \bibitemNoStop [0]{.\EOS\space}%
\providecommand \EOS [0]{\spacefactor3000\relax}%
\providecommand \BibitemShut  [1]{\csname bibitem#1\endcsname}%
\let\auto@bib@innerbib\@empty
\bibitem [{\citenamefont {Smith}\ \emph
  {et~al.}(2020{\natexlab{a}})\citenamefont {Smith}, \citenamefont {Nebgen},
  \citenamefont {Mathew}, \citenamefont {Chen}, \citenamefont {Lubbers},
  \citenamefont {Burakovsky}, \citenamefont {Tretiak}, \citenamefont {Nam},
  \citenamefont {Germann}, \citenamefont {Fensin},\ and\ \citenamefont
  {Barros}}]{Smith20}%
  \BibitemOpen
  \bibfield  {author} {\bibinfo {author} {\bibfnamefont {J.~S.}\ \bibnamefont
  {Smith}}, \bibinfo {author} {\bibfnamefont {B.}~\bibnamefont {Nebgen}},
  \bibinfo {author} {\bibfnamefont {N.}~\bibnamefont {Mathew}}, \bibinfo
  {author} {\bibfnamefont {J.}~\bibnamefont {Chen}}, \bibinfo {author}
  {\bibfnamefont {N.}~\bibnamefont {Lubbers}}, \bibinfo {author} {\bibfnamefont
  {L.}~\bibnamefont {Burakovsky}}, \bibinfo {author} {\bibfnamefont
  {S.}~\bibnamefont {Tretiak}}, \bibinfo {author} {\bibfnamefont {H.~A.}\
  \bibnamefont {Nam}}, \bibinfo {author} {\bibfnamefont {T.}~\bibnamefont
  {Germann}}, \bibinfo {author} {\bibfnamefont {S.}~\bibnamefont {Fensin}}, \
  and\ \bibinfo {author} {\bibfnamefont {K.}~\bibnamefont {Barros}},\
  }\href@noop {} {\  (\bibinfo {year} {2020}{\natexlab{a}})},\ \Eprint
  {http://arxiv.org/abs/2003.04934} {arXiv:2003.04934} \BibitemShut {NoStop}%
\bibitem [{\citenamefont {Lu}\ \emph {et~al.}(2020)\citenamefont {Lu},
  \citenamefont {Wang}, \citenamefont {Chen}, \citenamefont {Liu},
  \citenamefont {Lin}, \citenamefont {Car}, \citenamefont {E}, \citenamefont
  {Jia},\ and\ \citenamefont {Zhang}}]{Lu20}%
  \BibitemOpen
  \bibfield  {author} {\bibinfo {author} {\bibfnamefont {D.}~\bibnamefont
  {Lu}}, \bibinfo {author} {\bibfnamefont {H.}~\bibnamefont {Wang}}, \bibinfo
  {author} {\bibfnamefont {M.}~\bibnamefont {Chen}}, \bibinfo {author}
  {\bibfnamefont {J.}~\bibnamefont {Liu}}, \bibinfo {author} {\bibfnamefont
  {L.}~\bibnamefont {Lin}}, \bibinfo {author} {\bibfnamefont {R.}~\bibnamefont
  {Car}}, \bibinfo {author} {\bibfnamefont {W.}~\bibnamefont {E}}, \bibinfo
  {author} {\bibfnamefont {W.}~\bibnamefont {Jia}}, \ and\ \bibinfo {author}
  {\bibfnamefont {L.}~\bibnamefont {Zhang}},\ }\href@noop {} {\  (\bibinfo
  {year} {2020})},\ \Eprint {http://arxiv.org/abs/2004.11658}
  {arXiv:2004.11658} \BibitemShut {NoStop}%
\bibitem [{\citenamefont {Paszke}\ \emph {et~al.}(2019)\citenamefont {Paszke},
  \citenamefont {Gross}, \citenamefont {Massa}, \citenamefont {Lerer},
  \citenamefont {Bradbury}, \citenamefont {Chanan}, \citenamefont {Killeen},
  \citenamefont {Lin}, \citenamefont {Gimelshein}, \citenamefont {Antiga},
  \citenamefont {Desmaison}, \citenamefont {Kopf}, \citenamefont {Yang},
  \citenamefont {DeVito}, \citenamefont {Raison}, \citenamefont {Tejani},
  \citenamefont {Chilamkurthy}, \citenamefont {Steiner}, \citenamefont {Fang},
  \citenamefont {Bai},\ and\ \citenamefont {Chintala}}]{Paszke19}%
  \BibitemOpen
  \bibfield  {author} {\bibinfo {author} {\bibfnamefont {A.}~\bibnamefont
  {Paszke}}, \bibinfo {author} {\bibfnamefont {S.}~\bibnamefont {Gross}},
  \bibinfo {author} {\bibfnamefont {F.}~\bibnamefont {Massa}}, \bibinfo
  {author} {\bibfnamefont {A.}~\bibnamefont {Lerer}}, \bibinfo {author}
  {\bibfnamefont {J.}~\bibnamefont {Bradbury}}, \bibinfo {author}
  {\bibfnamefont {G.}~\bibnamefont {Chanan}}, \bibinfo {author} {\bibfnamefont
  {T.}~\bibnamefont {Killeen}}, \bibinfo {author} {\bibfnamefont
  {Z.}~\bibnamefont {Lin}}, \bibinfo {author} {\bibfnamefont {N.}~\bibnamefont
  {Gimelshein}}, \bibinfo {author} {\bibfnamefont {L.}~\bibnamefont {Antiga}},
  \bibinfo {author} {\bibfnamefont {A.}~\bibnamefont {Desmaison}}, \bibinfo
  {author} {\bibfnamefont {A.}~\bibnamefont {Kopf}}, \bibinfo {author}
  {\bibfnamefont {E.}~\bibnamefont {Yang}}, \bibinfo {author} {\bibfnamefont
  {Z.}~\bibnamefont {DeVito}}, \bibinfo {author} {\bibfnamefont
  {M.}~\bibnamefont {Raison}}, \bibinfo {author} {\bibfnamefont
  {A.}~\bibnamefont {Tejani}}, \bibinfo {author} {\bibfnamefont
  {S.}~\bibnamefont {Chilamkurthy}}, \bibinfo {author} {\bibfnamefont
  {B.}~\bibnamefont {Steiner}}, \bibinfo {author} {\bibfnamefont
  {L.}~\bibnamefont {Fang}}, \bibinfo {author} {\bibfnamefont {J.}~\bibnamefont
  {Bai}}, \ and\ \bibinfo {author} {\bibfnamefont {S.}~\bibnamefont
  {Chintala}},\ }in\ \href@noop {} {\emph {\bibinfo {booktitle} {Advances in
  {{Neural Information Processing Systems}} 32}}},\ \bibinfo {editor} {edited
  by\ \bibinfo {editor} {\bibfnamefont {H.}~\bibnamefont {Wallach}}, \bibinfo
  {editor} {\bibfnamefont {H.}~\bibnamefont {Larochelle}}, \bibinfo {editor}
  {\bibfnamefont {A.}~\bibnamefont {Beygelzimer}}, \bibinfo {editor}
  {\bibfnamefont {F.}~\bibnamefont {{d'Alch{\'e}-Buc}}}, \bibinfo {editor}
  {\bibfnamefont {E.}~\bibnamefont {Fox}}, \ and\ \bibinfo {editor}
  {\bibfnamefont {R.}~\bibnamefont {Garnett}}}\ (\bibinfo  {publisher} {{Curran
  Associates, Inc.}},\ \bibinfo {year} {2019})\ pp.\ \bibinfo {pages}
  {8026--8037}\BibitemShut {NoStop}%
\bibitem [{\citenamefont {Abadi}\ \emph {et~al.}(2016)\citenamefont {Abadi},
  \citenamefont {Agarwal}, \citenamefont {Barham}, \citenamefont {Brevdo},
  \citenamefont {Chen}, \citenamefont {Citro}, \citenamefont {Corrado},
  \citenamefont {Davis}, \citenamefont {Dean}, \citenamefont {Devin},
  \citenamefont {Ghemawat}, \citenamefont {Goodfellow}, \citenamefont {Harp},
  \citenamefont {Irving}, \citenamefont {Isard}, \citenamefont {Jia},
  \citenamefont {Jozefowicz}, \citenamefont {Kaiser}, \citenamefont {Kudlur},
  \citenamefont {Levenberg}, \citenamefont {Mane}, \citenamefont {Monga},
  \citenamefont {Moore}, \citenamefont {Murray}, \citenamefont {Olah},
  \citenamefont {Schuster}, \citenamefont {Shlens}, \citenamefont {Steiner},
  \citenamefont {Sutskever}, \citenamefont {Talwar}, \citenamefont {Tucker},
  \citenamefont {Vanhoucke}, \citenamefont {Vasudevan}, \citenamefont {Viegas},
  \citenamefont {Vinyals}, \citenamefont {Warden}, \citenamefont {Wattenberg},
  \citenamefont {Wicke}, \citenamefont {Yu},\ and\ \citenamefont
  {Zheng}}]{Abadi16}%
  \BibitemOpen
  \bibfield  {author} {\bibinfo {author} {\bibfnamefont {M.}~\bibnamefont
  {Abadi}}, \bibinfo {author} {\bibfnamefont {A.}~\bibnamefont {Agarwal}},
  \bibinfo {author} {\bibfnamefont {P.}~\bibnamefont {Barham}}, \bibinfo
  {author} {\bibfnamefont {E.}~\bibnamefont {Brevdo}}, \bibinfo {author}
  {\bibfnamefont {Z.}~\bibnamefont {Chen}}, \bibinfo {author} {\bibfnamefont
  {C.}~\bibnamefont {Citro}}, \bibinfo {author} {\bibfnamefont {G.~S.}\
  \bibnamefont {Corrado}}, \bibinfo {author} {\bibfnamefont {A.}~\bibnamefont
  {Davis}}, \bibinfo {author} {\bibfnamefont {J.}~\bibnamefont {Dean}},
  \bibinfo {author} {\bibfnamefont {M.}~\bibnamefont {Devin}}, \bibinfo
  {author} {\bibfnamefont {S.}~\bibnamefont {Ghemawat}}, \bibinfo {author}
  {\bibfnamefont {I.}~\bibnamefont {Goodfellow}}, \bibinfo {author}
  {\bibfnamefont {A.}~\bibnamefont {Harp}}, \bibinfo {author} {\bibfnamefont
  {G.}~\bibnamefont {Irving}}, \bibinfo {author} {\bibfnamefont
  {M.}~\bibnamefont {Isard}}, \bibinfo {author} {\bibfnamefont
  {Y.}~\bibnamefont {Jia}}, \bibinfo {author} {\bibfnamefont {R.}~\bibnamefont
  {Jozefowicz}}, \bibinfo {author} {\bibfnamefont {L.}~\bibnamefont {Kaiser}},
  \bibinfo {author} {\bibfnamefont {M.}~\bibnamefont {Kudlur}}, \bibinfo
  {author} {\bibfnamefont {J.}~\bibnamefont {Levenberg}}, \bibinfo {author}
  {\bibfnamefont {D.}~\bibnamefont {Mane}}, \bibinfo {author} {\bibfnamefont
  {R.}~\bibnamefont {Monga}}, \bibinfo {author} {\bibfnamefont
  {S.}~\bibnamefont {Moore}}, \bibinfo {author} {\bibfnamefont
  {D.}~\bibnamefont {Murray}}, \bibinfo {author} {\bibfnamefont
  {C.}~\bibnamefont {Olah}}, \bibinfo {author} {\bibfnamefont {M.}~\bibnamefont
  {Schuster}}, \bibinfo {author} {\bibfnamefont {J.}~\bibnamefont {Shlens}},
  \bibinfo {author} {\bibfnamefont {B.}~\bibnamefont {Steiner}}, \bibinfo
  {author} {\bibfnamefont {I.}~\bibnamefont {Sutskever}}, \bibinfo {author}
  {\bibfnamefont {K.}~\bibnamefont {Talwar}}, \bibinfo {author} {\bibfnamefont
  {P.}~\bibnamefont {Tucker}}, \bibinfo {author} {\bibfnamefont
  {V.}~\bibnamefont {Vanhoucke}}, \bibinfo {author} {\bibfnamefont
  {V.}~\bibnamefont {Vasudevan}}, \bibinfo {author} {\bibfnamefont
  {F.}~\bibnamefont {Viegas}}, \bibinfo {author} {\bibfnamefont
  {O.}~\bibnamefont {Vinyals}}, \bibinfo {author} {\bibfnamefont
  {P.}~\bibnamefont {Warden}}, \bibinfo {author} {\bibfnamefont
  {M.}~\bibnamefont {Wattenberg}}, \bibinfo {author} {\bibfnamefont
  {M.}~\bibnamefont {Wicke}}, \bibinfo {author} {\bibfnamefont
  {Y.}~\bibnamefont {Yu}}, \ and\ \bibinfo {author} {\bibfnamefont
  {X.}~\bibnamefont {Zheng}},\ }\href@noop {} {\  (\bibinfo {year} {2016})},\
  \Eprint {http://arxiv.org/abs/1603.04467} {arXiv:1603.04467} \BibitemShut
  {NoStop}%
\bibitem [{\citenamefont {Griewank}(1989)}]{Griewank89}%
  \BibitemOpen
  \bibfield  {author} {\bibinfo {author} {\bibfnamefont {A.}~\bibnamefont
  {Griewank}},\ }in\ \href@noop {} {\emph {\bibinfo {booktitle} {Mathematical
  {{Programming}}: {{Recent Developments}} and {{Applications}}}}},\ \bibinfo
  {editor} {edited by\ \bibinfo {editor} {\bibfnamefont {M.}~\bibnamefont
  {Iri}}\ and\ \bibinfo {editor} {\bibfnamefont {K.}~\bibnamefont {Tanabe}}}\
  (\bibinfo  {publisher} {{Kluwer Academic}},\ \bibinfo {address} {{Dordrecht,
  The Netherlands}},\ \bibinfo {year} {1989})\ pp.\ \bibinfo {pages}
  {83--108}\BibitemShut {NoStop}%
\bibitem [{\citenamefont {Wang}\ \emph
  {et~al.}(2018{\natexlab{a}})\citenamefont {Wang}, \citenamefont {Chern},
  \citenamefont {Batista},\ and\ \citenamefont {Barros}}]{Wang18b}%
  \BibitemOpen
  \bibfield  {author} {\bibinfo {author} {\bibfnamefont {Z.}~\bibnamefont
  {Wang}}, \bibinfo {author} {\bibfnamefont {G.-W.}\ \bibnamefont {Chern}},
  \bibinfo {author} {\bibfnamefont {C.~D.}\ \bibnamefont {Batista}}, \ and\
  \bibinfo {author} {\bibfnamefont {K.}~\bibnamefont {Barros}},\ }\href
  {\doibase 10.1063/1.5017741} {\bibfield  {journal} {\bibinfo  {journal} {J.
  Chem. Phys.}\ }\textbf {\bibinfo {volume} {148}},\ \bibinfo {pages} {094107}
  (\bibinfo {year} {2018}{\natexlab{a}})}\BibitemShut {NoStop}%
\bibitem [{\citenamefont {Bradbury}\ \emph {et~al.}(2018)\citenamefont
  {Bradbury}, \citenamefont {Frostig}, \citenamefont {Hawkins}, \citenamefont
  {Johnson}, \citenamefont {Leary}, \citenamefont {Maclaurin},\ and\
  \citenamefont {{Wanderman-Milne}}}]{Bradbury18}%
  \BibitemOpen
  \bibfield  {author} {\bibinfo {author} {\bibfnamefont {J.}~\bibnamefont
  {Bradbury}}, \bibinfo {author} {\bibfnamefont {R.}~\bibnamefont {Frostig}},
  \bibinfo {author} {\bibfnamefont {P.}~\bibnamefont {Hawkins}}, \bibinfo
  {author} {\bibfnamefont {M.~J.}\ \bibnamefont {Johnson}}, \bibinfo {author}
  {\bibfnamefont {C.}~\bibnamefont {Leary}}, \bibinfo {author} {\bibfnamefont
  {D.}~\bibnamefont {Maclaurin}}, \ and\ \bibinfo {author} {\bibfnamefont
  {S.}~\bibnamefont {{Wanderman-Milne}}},\ }\href@noop {} {\enquote {\bibinfo
  {title} {{{JAX}}: Composable transformations of {{Python}}+{{NumPy}}
  programs},}\ } (\bibinfo {year} {2018})\BibitemShut {NoStop}%
\bibitem [{\citenamefont {Schoenholz}\ and\ \citenamefont
  {Cubuk}(2019)}]{Schoenholz19}%
  \BibitemOpen
  \bibfield  {author} {\bibinfo {author} {\bibfnamefont {S.~S.}\ \bibnamefont
  {Schoenholz}}\ and\ \bibinfo {author} {\bibfnamefont {E.~D.}\ \bibnamefont
  {Cubuk}},\ }\href@noop {} {\  (\bibinfo {year} {2019})},\ \Eprint
  {http://arxiv.org/abs/1912.04232} {arXiv:1912.04232} \BibitemShut {NoStop}%
\bibitem [{\citenamefont {Innes}(2019)}]{Innes19}%
  \BibitemOpen
  \bibfield  {author} {\bibinfo {author} {\bibfnamefont {M.}~\bibnamefont
  {Innes}},\ }\href@noop {} {\  (\bibinfo {year} {2019})},\ \Eprint
  {http://arxiv.org/abs/1810.07951} {arXiv:1810.07951} \BibitemShut {NoStop}%
\bibitem [{\citenamefont {Artrith}\ and\ \citenamefont
  {Urban}(2016)}]{Artrith16}%
  \BibitemOpen
  \bibfield  {author} {\bibinfo {author} {\bibfnamefont {N.}~\bibnamefont
  {Artrith}}\ and\ \bibinfo {author} {\bibfnamefont {A.}~\bibnamefont
  {Urban}},\ }\href {\doibase 10.1016/j.commatsci.2015.11.047} {\bibfield
  {journal} {\bibinfo  {journal} {Comput. Mater. Sci.}\ }\textbf {\bibinfo
  {volume} {114}},\ \bibinfo {pages} {135} (\bibinfo {year}
  {2016})}\BibitemShut {NoStop}%
\bibitem [{\citenamefont {Singraber}(2018)}]{Singraber18}%
  \BibitemOpen
  \bibfield  {author} {\bibinfo {author} {\bibfnamefont {A.}~\bibnamefont
  {Singraber}},\ }\href@noop {} {} (\bibinfo {year} {2018}),\ \bibinfo {note}
  {{NP2P neural network potential, Available online,
  \url{https://compphysvienna.github.io/n2p2/}}}\BibitemShut {NoStop}%
\bibitem [{\citenamefont {Singraber}\ \emph {et~al.}(2019)\citenamefont
  {Singraber}, \citenamefont {Morawietz}, \citenamefont {Behler},\ and\
  \citenamefont {Dellago}}]{Singraber19}%
  \BibitemOpen
  \bibfield  {author} {\bibinfo {author} {\bibfnamefont {A.}~\bibnamefont
  {Singraber}}, \bibinfo {author} {\bibfnamefont {T.}~\bibnamefont
  {Morawietz}}, \bibinfo {author} {\bibfnamefont {J.}~\bibnamefont {Behler}}, \
  and\ \bibinfo {author} {\bibfnamefont {C.}~\bibnamefont {Dellago}},\ }\href
  {\doibase 10.1021/acs.jctc.8b01092} {\bibfield  {journal} {\bibinfo
  {journal} {J. Chem. Theory Comput.}\ }\textbf {\bibinfo {volume} {15}},\
  \bibinfo {pages} {3075} (\bibinfo {year} {2019})}\BibitemShut {NoStop}%
\bibitem [{\citenamefont {Smith}\ \emph {et~al.}(2017)\citenamefont {Smith},
  \citenamefont {Isayev},\ and\ \citenamefont {Roitberg}}]{Smith17}%
  \BibitemOpen
  \bibfield  {author} {\bibinfo {author} {\bibfnamefont {J.~S.}\ \bibnamefont
  {Smith}}, \bibinfo {author} {\bibfnamefont {O.}~\bibnamefont {Isayev}}, \
  and\ \bibinfo {author} {\bibfnamefont {A.~E.}\ \bibnamefont {Roitberg}},\
  }\href {\doibase 10.1039/C6SC05720A} {\bibfield  {journal} {\bibinfo
  {journal} {Chem. Sci.}\ }\textbf {\bibinfo {volume} {8}},\ \bibinfo {pages}
  {3192} (\bibinfo {year} {2017})}\BibitemShut {NoStop}%
\bibitem [{\citenamefont {Pun}\ \emph {et~al.}(2019)\citenamefont {Pun},
  \citenamefont {Batra}, \citenamefont {Ramprasad},\ and\ \citenamefont
  {Mishin}}]{Pun19}%
  \BibitemOpen
  \bibfield  {author} {\bibinfo {author} {\bibfnamefont {G.~P.~P.}\
  \bibnamefont {Pun}}, \bibinfo {author} {\bibfnamefont {R.}~\bibnamefont
  {Batra}}, \bibinfo {author} {\bibfnamefont {R.}~\bibnamefont {Ramprasad}}, \
  and\ \bibinfo {author} {\bibfnamefont {Y.}~\bibnamefont {Mishin}},\ }\href
  {\doibase 10.1038/s41467-019-10343-5} {\bibfield  {journal} {\bibinfo
  {journal} {Nat. Commun.}\ }\textbf {\bibinfo {volume} {10}},\ \bibinfo
  {pages} {2339} (\bibinfo {year} {2019})}\BibitemShut {NoStop}%
\bibitem [{\citenamefont {Lubbers}\ \emph {et~al.}(2018)\citenamefont
  {Lubbers}, \citenamefont {Smith},\ and\ \citenamefont {Barros}}]{Lubbers18}%
  \BibitemOpen
  \bibfield  {author} {\bibinfo {author} {\bibfnamefont {N.}~\bibnamefont
  {Lubbers}}, \bibinfo {author} {\bibfnamefont {J.~S.}\ \bibnamefont {Smith}},
  \ and\ \bibinfo {author} {\bibfnamefont {K.}~\bibnamefont {Barros}},\ }\href
  {\doibase 10.1063/1.5011181} {\bibfield  {journal} {\bibinfo  {journal} {J.
  Chem. Phys.}\ }\textbf {\bibinfo {volume} {148}},\ \bibinfo {pages} {241715}
  (\bibinfo {year} {2018})}\BibitemShut {NoStop}%
\bibitem [{\citenamefont {Sch{\"u}tt}\ \emph {et~al.}(2018)\citenamefont
  {Sch{\"u}tt}, \citenamefont {Sauceda}, \citenamefont {Kindermans},
  \citenamefont {Tkatchenko},\ and\ \citenamefont {M{\"u}ller}}]{Schutt18}%
  \BibitemOpen
  \bibfield  {author} {\bibinfo {author} {\bibfnamefont {K.~T.}\ \bibnamefont
  {Sch{\"u}tt}}, \bibinfo {author} {\bibfnamefont {H.~E.}\ \bibnamefont
  {Sauceda}}, \bibinfo {author} {\bibfnamefont {P.-J.}\ \bibnamefont
  {Kindermans}}, \bibinfo {author} {\bibfnamefont {A.}~\bibnamefont
  {Tkatchenko}}, \ and\ \bibinfo {author} {\bibfnamefont {K.-R.}\ \bibnamefont
  {M{\"u}ller}},\ }\href {\doibase 10.1063/1.5019779} {\bibfield  {journal}
  {\bibinfo  {journal} {J. Chem. Phys.}\ }\textbf {\bibinfo {volume} {148}},\
  \bibinfo {pages} {241722} (\bibinfo {year} {2018})}\BibitemShut {NoStop}%
\bibitem [{\citenamefont {Wang}\ \emph
  {et~al.}(2018{\natexlab{b}})\citenamefont {Wang}, \citenamefont {Zhang},
  \citenamefont {Han},\ and\ \citenamefont {E}}]{Wang18a}%
  \BibitemOpen
  \bibfield  {author} {\bibinfo {author} {\bibfnamefont {H.}~\bibnamefont
  {Wang}}, \bibinfo {author} {\bibfnamefont {L.}~\bibnamefont {Zhang}},
  \bibinfo {author} {\bibfnamefont {J.}~\bibnamefont {Han}}, \ and\ \bibinfo
  {author} {\bibfnamefont {W.}~\bibnamefont {E}},\ }\href {\doibase
  10.1016/j.cpc.2018.03.016} {\bibfield  {journal} {\bibinfo  {journal}
  {Comput. Phys. Commun.}\ }\textbf {\bibinfo {volume} {228}},\ \bibinfo
  {pages} {178} (\bibinfo {year} {2018}{\natexlab{b}})},\ \Eprint
  {http://arxiv.org/abs/1712.03641} {arXiv:1712.03641} \BibitemShut {NoStop}%
\bibitem [{\citenamefont {Yao}\ \emph {et~al.}(2018)\citenamefont {Yao},
  \citenamefont {Herr}, \citenamefont {Toth}, \citenamefont {Mckintyre},\ and\
  \citenamefont {Parkhill}}]{Yao18}%
  \BibitemOpen
  \bibfield  {author} {\bibinfo {author} {\bibfnamefont {K.}~\bibnamefont
  {Yao}}, \bibinfo {author} {\bibfnamefont {J.~E.}\ \bibnamefont {Herr}},
  \bibinfo {author} {\bibfnamefont {D.~W.}\ \bibnamefont {Toth}}, \bibinfo
  {author} {\bibfnamefont {R.}~\bibnamefont {Mckintyre}}, \ and\ \bibinfo
  {author} {\bibfnamefont {J.}~\bibnamefont {Parkhill}},\ }\href {\doibase
  10.1039/C7SC04934J} {\bibfield  {journal} {\bibinfo  {journal} {Chem. Sci.}\
  }\textbf {\bibinfo {volume} {9}},\ \bibinfo {pages} {2261} (\bibinfo {year}
  {2018})}\BibitemShut {NoStop}%
\bibitem [{\citenamefont {Zubatyuk}\ \emph {et~al.}(2019)\citenamefont
  {Zubatyuk}, \citenamefont {Smith}, \citenamefont {Leszczynski},\ and\
  \citenamefont {Isayev}}]{Zubatyuk19}%
  \BibitemOpen
  \bibfield  {author} {\bibinfo {author} {\bibfnamefont {R.}~\bibnamefont
  {Zubatyuk}}, \bibinfo {author} {\bibfnamefont {J.~S.}\ \bibnamefont {Smith}},
  \bibinfo {author} {\bibfnamefont {J.}~\bibnamefont {Leszczynski}}, \ and\
  \bibinfo {author} {\bibfnamefont {O.}~\bibnamefont {Isayev}},\ }\href
  {\doibase 10.1126/sciadv.aav6490} {\bibfield  {journal} {\bibinfo  {journal}
  {Sci. Adv.}\ }\textbf {\bibinfo {volume} {5}},\ \bibinfo {pages} {eaav6490}
  (\bibinfo {year} {2019})}\BibitemShut {NoStop}%
\bibitem [{\citenamefont {Unke}\ and\ \citenamefont {Meuwly}(2019)}]{Unke19}%
  \BibitemOpen
  \bibfield  {author} {\bibinfo {author} {\bibfnamefont {O.~T.}\ \bibnamefont
  {Unke}}\ and\ \bibinfo {author} {\bibfnamefont {M.}~\bibnamefont {Meuwly}},\
  }\href {\doibase 10.1021/acs.jctc.9b00181} {\bibfield  {journal} {\bibinfo
  {journal} {J. Chem. Theory Comput.}\ }\textbf {\bibinfo {volume} {15}},\
  \bibinfo {pages} {3678} (\bibinfo {year} {2019})}\BibitemShut {NoStop}%
\bibitem [{\citenamefont {Gao}\ \emph {et~al.}(2020)\citenamefont {Gao},
  \citenamefont {Ramezanghorbani}, \citenamefont {Isayev}, \citenamefont
  {Smith},\ and\ \citenamefont {Roitberg}}]{Gao20}%
  \BibitemOpen
  \bibfield  {author} {\bibinfo {author} {\bibfnamefont {X.}~\bibnamefont
  {Gao}}, \bibinfo {author} {\bibfnamefont {F.}~\bibnamefont
  {Ramezanghorbani}}, \bibinfo {author} {\bibfnamefont {O.}~\bibnamefont
  {Isayev}}, \bibinfo {author} {\bibfnamefont {J.}~\bibnamefont {Smith}}, \
  and\ \bibinfo {author} {\bibfnamefont {A.}~\bibnamefont {Roitberg}},\ }\href
  {\doibase 10.26434/chemrxiv.12218294.v1} {\  (\bibinfo {year} {2020}),\
  10.26434/chemrxiv.12218294.v1}\BibitemShut {NoStop}%
\bibitem [{\citenamefont {Gilmer}\ \emph {et~al.}(2020)\citenamefont {Gilmer},
  \citenamefont {Schoenholz}, \citenamefont {Riley}, \citenamefont {Vinyals},\
  and\ \citenamefont {Dahl}}]{Gilmer20}%
  \BibitemOpen
  \bibfield  {author} {\bibinfo {author} {\bibfnamefont {J.}~\bibnamefont
  {Gilmer}}, \bibinfo {author} {\bibfnamefont {S.~S.}\ \bibnamefont
  {Schoenholz}}, \bibinfo {author} {\bibfnamefont {P.~F.}\ \bibnamefont
  {Riley}}, \bibinfo {author} {\bibfnamefont {O.}~\bibnamefont {Vinyals}}, \
  and\ \bibinfo {author} {\bibfnamefont {G.~E.}\ \bibnamefont {Dahl}},\ }in\
  \href {\doibase 10.1007/978-3-030-40245-7_10} {\emph {\bibinfo {booktitle}
  {Machine {{Learning Meets Quantum Physics}}}}},\ \bibinfo {series and number}
  {Lecture {{Notes}} in {{Physics}}},\ \bibinfo {editor} {edited by\ \bibinfo
  {editor} {\bibfnamefont {K.~T.}\ \bibnamefont {Sch{\"u}tt}}, \bibinfo
  {editor} {\bibfnamefont {S.}~\bibnamefont {Chmiela}}, \bibinfo {editor}
  {\bibfnamefont {O.~A.}\ \bibnamefont {{von Lilienfeld}}}, \bibinfo {editor}
  {\bibfnamefont {A.}~\bibnamefont {Tkatchenko}}, \bibinfo {editor}
  {\bibfnamefont {K.}~\bibnamefont {Tsuda}}, \ and\ \bibinfo {editor}
  {\bibfnamefont {K.-R.}\ \bibnamefont {M{\"u}ller}}}\ (\bibinfo  {publisher}
  {{Springer International Publishing}},\ \bibinfo {address} {{Cham}},\
  \bibinfo {year} {2020})\ pp.\ \bibinfo {pages} {199--214}\BibitemShut
  {NoStop}%
\bibitem [{Note1()}]{Note1}%
  \BibitemOpen
  \bibinfo {note} {For efficiency, these gradients should be evaluated using
  backpropagation. Implementing a \protect \emph {first }iteration of
  backpropagation manually is not too difficult.}\BibitemShut {Stop}%
\bibitem [{\citenamefont {Pearlmutter}(1994)}]{Pearlmutter94}%
  \BibitemOpen
  \bibfield  {author} {\bibinfo {author} {\bibfnamefont {B.~A.}\ \bibnamefont
  {Pearlmutter}},\ }\href {\doibase 10.1162/neco.1994.6.1.147} {\bibfield
  {journal} {\bibinfo  {journal} {Neural Computation}\ }\textbf {\bibinfo
  {volume} {6}},\ \bibinfo {pages} {147} (\bibinfo {year} {1994})}\BibitemShut
  {NoStop}%
\bibitem [{\citenamefont {Trott}\ \emph {et~al.}(2014)\citenamefont {Trott},
  \citenamefont {Hammond},\ and\ \citenamefont {Thompson}}]{Trott14}%
  \BibitemOpen
  \bibfield  {author} {\bibinfo {author} {\bibfnamefont {C.~R.}\ \bibnamefont
  {Trott}}, \bibinfo {author} {\bibfnamefont {S.~D.}\ \bibnamefont {Hammond}},
  \ and\ \bibinfo {author} {\bibfnamefont {A.~P.}\ \bibnamefont {Thompson}},\
  }in\ \href {\doibase 10.1007/978-3-319-07518-1_2} {\emph {\bibinfo
  {booktitle} {Supercomputing}}},\ \bibinfo {series and number} {Lecture
  {{Notes}} in {{Computer Science}}},\ \bibinfo {editor} {edited by\ \bibinfo
  {editor} {\bibfnamefont {J.~M.}\ \bibnamefont {Kunkel}}, \bibinfo {editor}
  {\bibfnamefont {T.}~\bibnamefont {Ludwig}}, \ and\ \bibinfo {editor}
  {\bibfnamefont {H.~W.}\ \bibnamefont {Meuer}}}\ (\bibinfo  {publisher}
  {{Springer International Publishing}},\ \bibinfo {address} {{Cham}},\
  \bibinfo {year} {2014})\ pp.\ \bibinfo {pages} {19--34}\BibitemShut {NoStop}%
\bibitem [{\citenamefont {Thompson}\ \emph {et~al.}(2015)\citenamefont
  {Thompson}, \citenamefont {Swiler}, \citenamefont {Trott}, \citenamefont
  {Foiles},\ and\ \citenamefont {Tucker}}]{Thompson15}%
  \BibitemOpen
  \bibfield  {author} {\bibinfo {author} {\bibfnamefont {A.~P.}\ \bibnamefont
  {Thompson}}, \bibinfo {author} {\bibfnamefont {L.~P.}\ \bibnamefont
  {Swiler}}, \bibinfo {author} {\bibfnamefont {C.~R.}\ \bibnamefont {Trott}},
  \bibinfo {author} {\bibfnamefont {S.~M.}\ \bibnamefont {Foiles}}, \ and\
  \bibinfo {author} {\bibfnamefont {G.~J.}\ \bibnamefont {Tucker}},\ }\href
  {\doibase 10.1016/j.jcp.2014.12.018} {\bibfield  {journal} {\bibinfo
  {journal} {J. Comput. Phys.}\ }\textbf {\bibinfo {volume} {285}},\ \bibinfo
  {pages} {316} (\bibinfo {year} {2015})}\BibitemShut {NoStop}%
\bibitem [{\citenamefont {He}\ \emph {et~al.}(2015)\citenamefont {He},
  \citenamefont {Zhang}, \citenamefont {Ren},\ and\ \citenamefont
  {Sun}}]{He15}%
  \BibitemOpen
  \bibfield  {author} {\bibinfo {author} {\bibfnamefont {K.}~\bibnamefont
  {He}}, \bibinfo {author} {\bibfnamefont {X.}~\bibnamefont {Zhang}}, \bibinfo
  {author} {\bibfnamefont {S.}~\bibnamefont {Ren}}, \ and\ \bibinfo {author}
  {\bibfnamefont {J.}~\bibnamefont {Sun}},\ }in\ \href {\doibase
  10.1109/ICCV.2015.123} {\emph {\bibinfo {booktitle} {Proceedings of the 2015
  {{IEEE International Conference}} on {{Computer Vision}} ({{ICCV}})}}},\
  \bibinfo {series and number} {{{ICCV}} '15}\ (\bibinfo  {publisher} {{IEEE
  Computer Society}},\ \bibinfo {address} {{USA}},\ \bibinfo {year} {2015})\
  pp.\ \bibinfo {pages} {1026--1034}\BibitemShut {NoStop}%
\bibitem [{\citenamefont {Behler}\ and\ \citenamefont
  {Parrinello}(2007)}]{Behler07}%
  \BibitemOpen
  \bibfield  {author} {\bibinfo {author} {\bibfnamefont {J.}~\bibnamefont
  {Behler}}\ and\ \bibinfo {author} {\bibfnamefont {M.}~\bibnamefont
  {Parrinello}},\ }\href {\doibase 10.1103/PhysRevLett.98.146401} {\bibfield
  {journal} {\bibinfo  {journal} {Phys. Rev. Lett.}\ }\textbf {\bibinfo
  {volume} {98}},\ \bibinfo {pages} {146401} (\bibinfo {year}
  {2007})}\BibitemShut {NoStop}%
\bibitem [{\citenamefont {Smith}(2020)}]{Smith20b}%
  \BibitemOpen
  \bibfield  {author} {\bibinfo {author} {\bibfnamefont {J.~S.}\ \bibnamefont
  {Smith}},\ }\href@noop {} {} (\bibinfo {year} {2020}),\ \bibinfo {note}
  {{Neurochem binaries, Available online,
  \url{https://github.com/isayev/ASE_ANI}}}\BibitemShut {NoStop}%
\bibitem [{\citenamefont {Smith}\ \emph
  {et~al.}(2020{\natexlab{b}})\citenamefont {Smith}, \citenamefont {Zubatyuk},
  \citenamefont {Nebgen}, \citenamefont {Lubbers}, \citenamefont {Barros},
  \citenamefont {Roitberg}, \citenamefont {Isayev},\ and\ \citenamefont
  {Tretiak}}]{Smith20a}%
  \BibitemOpen
  \bibfield  {author} {\bibinfo {author} {\bibfnamefont {J.~S.}\ \bibnamefont
  {Smith}}, \bibinfo {author} {\bibfnamefont {R.}~\bibnamefont {Zubatyuk}},
  \bibinfo {author} {\bibfnamefont {B.}~\bibnamefont {Nebgen}}, \bibinfo
  {author} {\bibfnamefont {N.}~\bibnamefont {Lubbers}}, \bibinfo {author}
  {\bibfnamefont {K.}~\bibnamefont {Barros}}, \bibinfo {author} {\bibfnamefont
  {A.~E.}\ \bibnamefont {Roitberg}}, \bibinfo {author} {\bibfnamefont
  {O.}~\bibnamefont {Isayev}}, \ and\ \bibinfo {author} {\bibfnamefont
  {S.}~\bibnamefont {Tretiak}},\ }\href {\doibase 10.1038/s41597-020-0473-z}
  {\bibfield  {journal} {\bibinfo  {journal} {Sci. Data}\ }\textbf {\bibinfo
  {volume} {7}},\ \bibinfo {pages} {134} (\bibinfo {year}
  {2020}{\natexlab{b}})}\BibitemShut {NoStop}%
\bibitem [{\citenamefont {Smith}\ \emph {et~al.}(2018)\citenamefont {Smith},
  \citenamefont {Nebgen}, \citenamefont {Lubbers}, \citenamefont {Isayev},\
  and\ \citenamefont {Roitberg}}]{Smith18}%
  \BibitemOpen
  \bibfield  {author} {\bibinfo {author} {\bibfnamefont {J.~S.}\ \bibnamefont
  {Smith}}, \bibinfo {author} {\bibfnamefont {B.}~\bibnamefont {Nebgen}},
  \bibinfo {author} {\bibfnamefont {N.}~\bibnamefont {Lubbers}}, \bibinfo
  {author} {\bibfnamefont {O.}~\bibnamefont {Isayev}}, \ and\ \bibinfo {author}
  {\bibfnamefont {A.~E.}\ \bibnamefont {Roitberg}},\ }\href {\doibase
  10.1063/1.5023802} {\bibfield  {journal} {\bibinfo  {journal} {J. Chem.
  Phys.}\ }\textbf {\bibinfo {volume} {148}},\ \bibinfo {pages} {241733}
  (\bibinfo {year} {2018})}\BibitemShut {NoStop}%
\bibitem [{\citenamefont {Cooper}\ \emph {et~al.}(2020)\citenamefont {Cooper},
  \citenamefont {K{\"a}stner}, \citenamefont {Urban},\ and\ \citenamefont
  {Artrith}}]{Cooper20}%
  \BibitemOpen
  \bibfield  {author} {\bibinfo {author} {\bibfnamefont {A.~M.}\ \bibnamefont
  {Cooper}}, \bibinfo {author} {\bibfnamefont {J.}~\bibnamefont {K{\"a}stner}},
  \bibinfo {author} {\bibfnamefont {A.}~\bibnamefont {Urban}}, \ and\ \bibinfo
  {author} {\bibfnamefont {N.}~\bibnamefont {Artrith}},\ }\href {\doibase
  10.1038/s41524-020-0323-8} {\bibfield  {journal} {\bibinfo  {journal} {npj
  Comput. Mater.}\ }\textbf {\bibinfo {volume} {6}},\ \bibinfo {pages} {1}
  (\bibinfo {year} {2020})}\BibitemShut {NoStop}%
\bibitem [{\citenamefont {Bart{\'o}k}\ and\ \citenamefont
  {Cs{\'a}nyi}(2015)}]{Bartok15}%
  \BibitemOpen
  \bibfield  {author} {\bibinfo {author} {\bibfnamefont {A.~P.}\ \bibnamefont
  {Bart{\'o}k}}\ and\ \bibinfo {author} {\bibfnamefont {G.}~\bibnamefont
  {Cs{\'a}nyi}},\ }\href {\doibase 10.1002/qua.24927} {\bibfield  {journal}
  {\bibinfo  {journal} {International Journal of Quantum Chemistry}\ }\textbf
  {\bibinfo {volume} {115}},\ \bibinfo {pages} {1051} (\bibinfo {year}
  {2015})}\BibitemShut {NoStop}%
\bibitem [{\citenamefont {Rupp}(2015)}]{Rupp15}%
  \BibitemOpen
  \bibfield  {author} {\bibinfo {author} {\bibfnamefont {M.}~\bibnamefont
  {Rupp}},\ }\href {\doibase 10.1002/qua.24954} {\bibfield  {journal} {\bibinfo
   {journal} {Int. J. Quantum Chem.}\ }\textbf {\bibinfo {volume} {115}},\
  \bibinfo {pages} {1058} (\bibinfo {year} {2015})}\BibitemShut {NoStop}%
\bibitem [{\citenamefont {Chmiela}\ \emph {et~al.}(2017)\citenamefont
  {Chmiela}, \citenamefont {Tkatchenko}, \citenamefont {Sauceda}, \citenamefont
  {Poltavsky}, \citenamefont {Sch{\"u}tt},\ and\ \citenamefont
  {M{\"u}ller}}]{Chmiela17}%
  \BibitemOpen
  \bibfield  {author} {\bibinfo {author} {\bibfnamefont {S.}~\bibnamefont
  {Chmiela}}, \bibinfo {author} {\bibfnamefont {A.}~\bibnamefont {Tkatchenko}},
  \bibinfo {author} {\bibfnamefont {H.~E.}\ \bibnamefont {Sauceda}}, \bibinfo
  {author} {\bibfnamefont {I.}~\bibnamefont {Poltavsky}}, \bibinfo {author}
  {\bibfnamefont {K.~T.}\ \bibnamefont {Sch{\"u}tt}}, \ and\ \bibinfo {author}
  {\bibfnamefont {K.-R.}\ \bibnamefont {M{\"u}ller}},\ }\href {\doibase
  10.1126/sciadv.1603015} {\bibfield  {journal} {\bibinfo  {journal} {Sci.
  Adv.}\ }\textbf {\bibinfo {volume} {3}},\ \bibinfo {pages} {e1603015}
  (\bibinfo {year} {2017})}\BibitemShut {NoStop}%
\bibitem [{\citenamefont {Christensen}\ \emph {et~al.}(2019)\citenamefont
  {Christensen}, \citenamefont {Faber},\ and\ \citenamefont {{von
  Lilienfeld}}}]{Christensen19}%
  \BibitemOpen
  \bibfield  {author} {\bibinfo {author} {\bibfnamefont {A.~S.}\ \bibnamefont
  {Christensen}}, \bibinfo {author} {\bibfnamefont {F.~A.}\ \bibnamefont
  {Faber}}, \ and\ \bibinfo {author} {\bibfnamefont {O.~A.}\ \bibnamefont {{von
  Lilienfeld}}},\ }\href {\doibase 10.1063/1.5053562} {\bibfield  {journal}
  {\bibinfo  {journal} {J. Chem. Phys.}\ }\textbf {\bibinfo {volume} {150}},\
  \bibinfo {pages} {064105} (\bibinfo {year} {2019})}\BibitemShut {NoStop}%
\end{thebibliography}%

\end{document}